# White Paper
*(Submitted to ExoPlanet Task Force)*

# Terrestrial and Habitable Planet Formation in Binary and Multi-star Systems

## Authors

Nader Haghighipour
*Institute for Astronomy and NASA Astrobiology Institute, University of Hawaii-Manoa*

Steinn Sigurdsson
*Department of Astronomy & Astrophysics and NASA Astrobiology Institute, Penn State University*

Jack Lissauer
*Space Science and Astrobiology Division, NASA-Ames Research Center*

Sean Raymond
*Center for Astrophysics and Space Astronomy, and Center for Astrobiology, University of Colorado*

## Introduction

The discovery of extrasolar planets during the past decade has confronted astronomers with many new challenges. The diverse and surprising dynamical characteristics of many of these objects have made scientists wonder to what extent the current theories of planet formation can be applied to other planetary systems. A major challenge of planetary science is now to explain how such planets were formed, how they acquired their unfamiliar dynamical state, whether there are habitable extrasolar planets, and how to detect such habitable worlds.

In this respect, one of the most surprising discoveries is the detection of planets in binary star systems. Among the currently known extrasolar planet-hosting stars approximately 25% are members of binaries (Table 1). With the exception of the pulsar planetary system PSR B1620-26 (Sigurdsson et al. 2003; Richer et al. 2003; Beer et al. 2004), and possibly the system of HD202206 (Correia et al. 2005), planets in these binary systems revolve only around one of the stars. While the majority of these binaries are wide (i.e., with separations between 250 and 6500 AU, where the perturbative effect of the stellar companion on planet formation around the other star is negligible), the detection of Jovian-type planets in the two binaries γ Cephei (separation of 18.5 AU, see Hatzes et al. 2003) and GJ 86 (separation of 21 AU, see Els et al. 2001) have brought to the forefront questions on the formation of giant planets and the possibility of the existence of smaller bodies in moderately close binary and multiple star systems. Given that more than half of main sequence stars are members of binaries/multiples (Duquennoy & Mayor 1991; Mathieu et al. 2000), and the frequency of planets in binary/multiple systems is comparable to those around single stars (Bonavita & Desidera 2007), such questions have realistic grounds.

At present, the sensitivity of the detection techniques does not allow routine discovery of Earth-sized objects around binary and multi-star systems. However, with the advancement of new techniques, and with the recent launch of CoRoT and the launch of Kepler in late 2008, the detection of more planets (possibly terrestrial-class objects) in such systems is on the horizon.

Table 1- Binary and multi-star systems with extrasolar planets (Haghighipour 2006)

| Star | Star | Star | Star |
| --- | --- | --- | --- |
| HD142 (GJ 9002) | HD3651 | HD9826 (υAnd) | HD13445 (GJ 86) |
| HD19994 | HD22049 (εEri) | HD27442 | HD40979 |
| HD41004 | HD75732 (55 Cnc) | HD80606 | HD89744 |
| HD114762 | HD117176 (70 Vir) | HD120136 (τBoo) | HD121504 |
| HD137759 | HD143761 (ρCrb) | HD178911 | HD186472 (16 Cyg) |
| HD190360 (GJ 777 A) | HD192263 | HD195019 | HD213240 |
| HD217107 | HD219449 | HD219542 | HD222404 (γCeph) |
| HD178911 | PSR B1257-20 | PSR B1620-26 | HD202206 |

See http://www.obspm.fr/planets for complete list of extrasolar planets with their corresponding references.

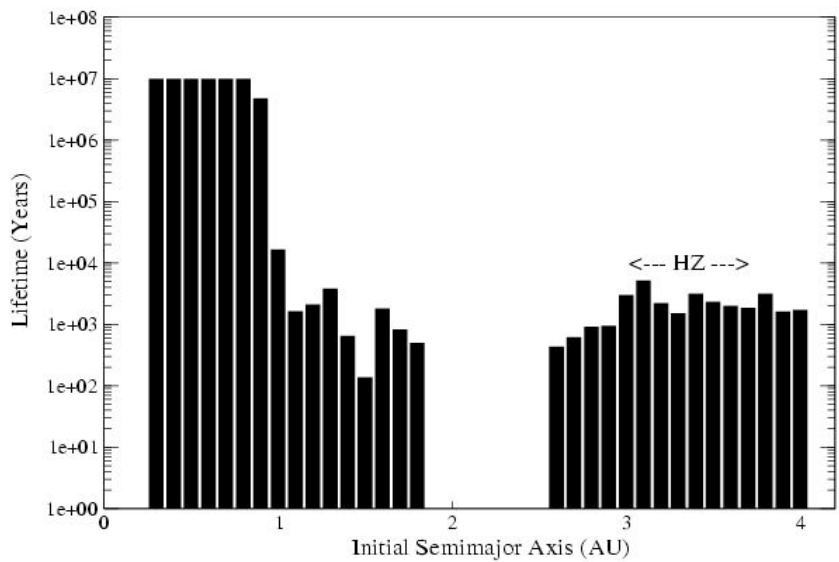

**Fig 1.** The time of ejection, vs. the initial semimajor axis of an Earth-like planet in a co-planar arrangement in the γ Cephei system. The binary consists of a 1.59 solar-masses K1 IV subgiant as its primary (Fuhrmann 2003) and a probable red M dwarf with a mass of 0.41 solar-masses (Neuhauser et al 2007) as its secondary. The semimajor axis and eccentricity of the binary are 18.5 AU and 0.36, respectively (Hatzes et al. 2003). The primary star is host to a 1.7 Jupiter-masses object in an orbit with a semimajor axis of 2.13 AU and eccentricity of 0.12. The habitable zone of the primary is within 3 AU to 3.7 AU from this star (Haghighipour 2006). As shown here, the orbit of an Earth-sized object in the primary's habitable zone is unstable. However, such an object can have a log-term stable orbit in distances closer to the primary star.

Theoretical studies and numerical modeling of terrestrial and habitable planet formation in such dynamically complex environments are, therefore, necessary to gain fundamental insights into the prospects for life in such systems and have great strategic impact on NASA science and missions.

Several lines of investigations are needed to ensure progress in understanding the formation of terrestrial and habitable planets in binary and multi-star systems.

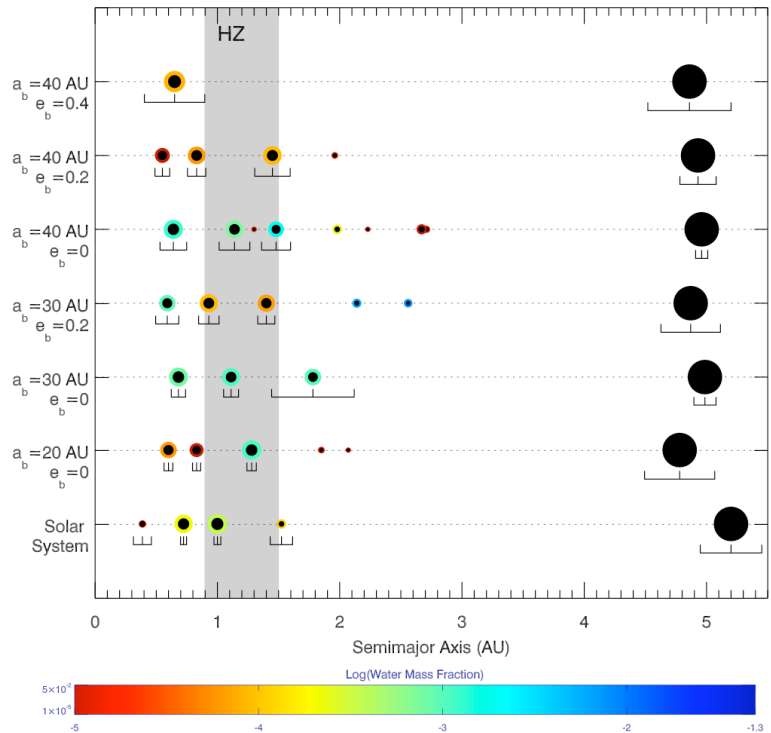

**Fig 2.** Results of simulations of the formation of Earth-like objects in the habitable zone of the primary of a binary star system. The stars of the binary are Sun-like and the primary is host to a Jupiter-sized object on a circular orbit at 5 AU. Simulations show the results for different values of the eccentricity and semimajor axis of the stellar companion (Haghighipour & Raymond 2007). As shown here, the orbital motion of the secondary star disturbs the orbit of the giant planet, which in turn affects the final assembly and water contents of the terrestrial objects. This figure also shows that binary systems with larger perihelia are more favorable for forming and harboring habitable planets. The quantities $a_b$ and $e_b$ represent the seimmajor axis and eccentricity of the binary.

## 1) Computational Modeling

Extensive numerical studies are necessary to

 i) map the parameter-space of binary and/or multiple star systems to identify regions where giant and terrestrial planets can have long-term stable orbits,
 ii) simulate the collision and growth of planetesimals to form protoplanetary objects,
 iii) simulate the formation of planetesimals in circumbinary and circumstellar disks,
 iii) develop models of protoplanet disk chemistry that ensure delivery of water to terrestrial-class planets in the habitable zone,
 iv) simulate the interaction of planetary embryos and the late stage of terrestrial planet formation.

The parameter-space is large and includes the masses and orbital parameters of the stars and planets. It is, therefore, necessary to develop a systematic approach, based on the results of current research, to avoid un-necessary simulations, particularly if the computational resources are limited.

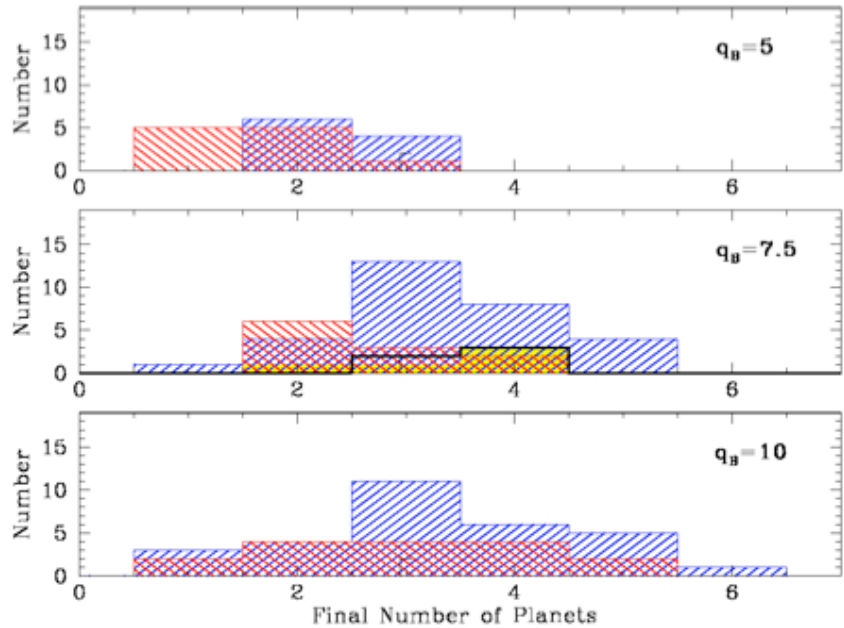

**Fig 3.** Histograms of the number of final terrestrial planets formed in binary star systems with periastron distances of 5 AU (top), 7.5 AU (middle), and 10 AU (bottom). The color red corresponds to simulations in a binary in which the primary and secondary stars are 0.5 solar-masses. The color blue represents a binary with 1 solar-mass stars, and the color yellow corresponds to a binary with a 0.5 solar-masses primary and a 1 solar-mass secondary. The black line in the middle panel shows the results of simulations when the primary star is 1 solar-mass and the secondary is 0.5 solar-masses. As shown here, the typical number of final planets clearly increases in systems with larger stellar periastra, and also when the companion star is less massive than the primary (Quintana et al. 2007).

Current research has indicated that terrestrial-class planets can have long-term stable orbits as long as they are closer to their host stars and their orbits lie outside the *influence zone* of the giant planet of the system (figure 1, also see Holman & Wiegert 1999; David et al. 2003; Haghighipour 2006). This implies, in order for such systems to be habitable, the habitable zone of the planet-hosting star has to be considerably closer to it than orbit of its giant planet(s). Given that the location of the habitable zone is a function of the luminosity of a star, the above-mentioned criterion can be used to constrain stellar properties. Recent numerical simulations have also shown that (1) water-delivery is more efficient when the perihelion of the binary is large and the orbit of the giant planet is close to a circle (figures 2 and 3, also see Quintana et al. 2007, Haghighipour & Raymond 2007), and (2) habitable planets can form in the habitable zone of a star during the migration of giant planets (figure 4, see Raymond, Mandell & Sigurdsson, 2006). Since many stars are formed in clusters, their mutual interactions may change their orbital configurations and cause their giant planets to revolve around their host starts in un-conventional orbits. Theoretical studies are essential to identify systems capable of forming and harboring habitable planets.

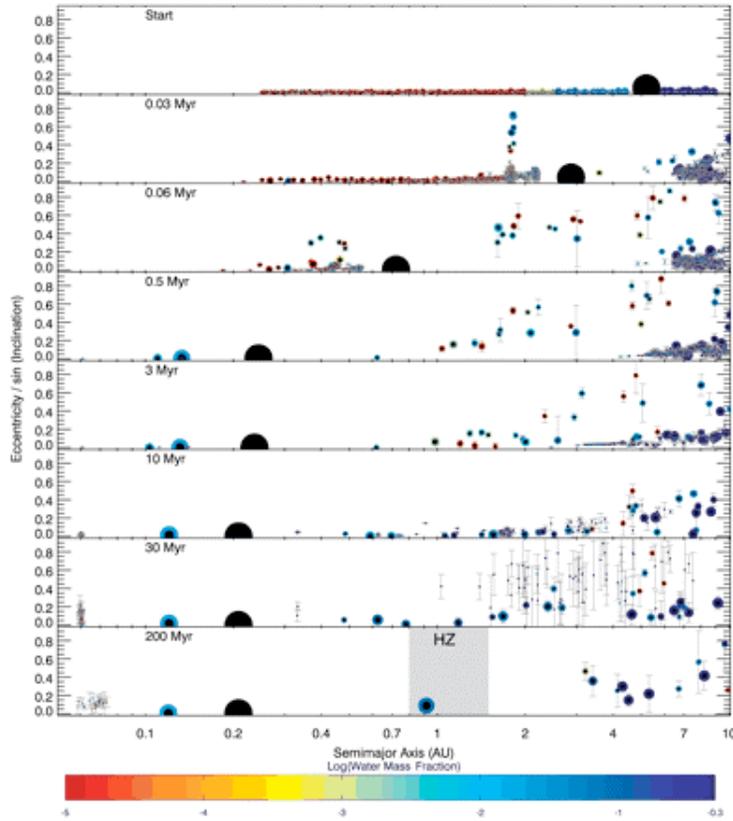

**Fig 4.** Habitable planet formation at presence of giant planet migration (Raymond, Mandell & Sigurdsson 2006). The system consists of a Sun-like star and a Jupiter-sized giant planet. The figure shows snapshots in time of the evolution of one simulation. Each panel plots the orbital eccentricity versus semimajor axis for each surviving body. The size of each body is proportional to its physical size (except for the giant planet, shown in black). The vertical "error bars" represent the sine of each body's inclination on the *y*-axis scale. The color of each dot corresponds to its water content (as per the color bar), and the dark inner dot represents the relative size of its iron core. For scale, the Earth's water content is roughly $10^{-3}$. As shown here, an Earth-like object can form in the habitable zone of the star while the giant planet migrates to closer distances.

**2) Theoretical Analysis of Observation Data**

Recent observations of binary star systems, using Spitzer Space Telescope, show evidence of debris disks in these environments (Trilling et al. 2007). As shown by these authors, approximately 60% of their observed close binary systems (separation smaller than 3 AU) have excess in their thermal emissions, implying on-going collisions in their planetesimal regions. Future space-, air-, and ground-based telescopes such as ALMA, SOFIA and JWST will be able to detect more of such disks and will also be able to resolve their fine structures. Numerical simulations, similar to those for debris disks around single stars (Telesco et al. 2005), will be necessary in order to understand the dynamics of such planet-forming environments, and also identify the source of their disks features (e.g., embedded planets, and/or on-going planetesimals collision). Due to the complex nature of these systems, such numerical studies require more advanced computational codes, and more powerful computers. Developing theories of disk evolution in close binary systems is also essential.

### 3) Computational Resources

Given the extent and complexity of simulations of planet formation in multi-star systems, and the high dimensionality of the parameter space of initial conditions, supports for developing computational resources with the primary focus of conducting numerical analysis of terrestrial planet formation are essential. Reliable simulations of collisional growth of planetesimals and planetary embryos require integration of the orbits of several hundred thousands of such objects. With the current technology, such simulations may take several months to a year to complete. It is therefore necessary to develop (i) faster integration routines, and (ii) major computational facilities with the primary focus of simulating terrestrial planet formation.

## Strategic Impact to NASA Missions

Understanding terrestrial and habitable planet formation in binary and multiple star systems has implications for investigating the habitability of extra-solar planets. It ties directly into near future NASA missions, in particular Kepler, and JWST as well as complementary ongoing and planned NSF and privately funded surveys that include transit, and radial velocity. It is also closely coupled with the scientific aspect of the Space Exploration Vision and aligns with the 2006 NASA Science Program implementation of the Strategic sub-goal 3D:

*"Discover the origin, structure, evolution, and destiny of the universe and search for earth-like planets."*

The strategic relevance to the NASA missions is in the prospects for detection of habitable Earth-like planets. Studies such as those presented here underlie hypotheses regarding the likelihood of the existence of such planets, the origin of life in the habitable zones of their host stars, and theories of evolution and persistence of life after initiation, at the presence of a stellar companion. Earth-like objects in and around binary star systems allow testing of theories of extrasolar habitability and origin of life. Prospects for testing of extrasolar life are intrinsically exciting and valuable to the NASA community and the public, and the systems to be explored, once found, provide calibration targets for future NASA missions.